\documentclass[letterpaper,twocolumn,prl,reprint,amsmath]{revtex4-2}
\usepackage[latin9]{inputenc}
\usepackage{amsmath}
\usepackage{amssymb}
\usepackage{graphicx}

\makeatletter

\pdfpageheight\paperheight
\pdfpagewidth\paperwidth

\usepackage{color}
\usepackage{float}
\makeatother

\begin{document}
\maketitle \textbf{Comment on ``Superconductivity and
Mott Physics in Organic Charge Transfer Materials"} \vskip 1pc

\noindent Menke \textit{et al.} \cite{Menke24a} recently claimed that
superconductivity (SC) in the $\kappa$-phase organic charge-transfer
solids (CTS) can be understood within the two-dimensional half-filled
anisotropic triangular-lattice Hubbard model.  Experimentally,
$\kappa$-CTS are mostly but not always antiferromagnetic (AFM) at
ambient pressure and SC appears under pressure. In apparent agreement
with this observation, Menke \textit{et al.} found AFM ground states
for small $t/U$ and SC over a small region at the interface of AFM and
Fermi liquid ground states with increasing $t/U$ at fixed $t'/t$,
where $U$ is the Hubbard repulsion. Menke \textit{et al's}
computational results directly contradict those obtained using exact
diagonalization \cite{Clay08a} and Path Integral Renormalization Group
(PIRG) \cite{Dayal12a} approaches. It is clearly of interest to
determine the origin of this discrepancy, especially in view of the
facts that (a) related arguments continue to persist in the context of
cuprate SC superconductivity (which however involves doping), and (b)
there exist CTS in which SC is not proximate to AFM, but is separated
by an intermediate charge-disproportionated phase
\cite{Hashimoto15a}. Here we show that Menke \textit{et al's}
conclusion regarding SC is incorrect and originates from a flawed
assumption.

Menke \textit{et al.} established the superconducting ground state
using Cluster Dynamic Mean Field Theory (CDMFT), within which
momentum-dependent susceptibilities are calculated by summing over all
sites of a seven-site cluster. This approach does not distinguish
between Cooper pairs separated by short versus long distances.  We
have performed computations on the triangular lattice Hubbard
Hamiltonian for 4$\times$4, 6$\times$4, and 6$\times$6 lattices using
exact diagonalization and PIRG \cite{Imada00a,Mizusaki04a} for the
same $t^{\prime}=0.4t$, $t=1$ as Menke \textit{et al.} Fig.~1 (for the
6$\times$4 lattice the ground state occurs in the total spin $S=1$
subspace; our calculations are for the lowest $S=0$ state).  For these
clusters PIRG is essentially exact \cite{Imada00a,Mizusaki04a}.  With
our choice of axes, $d_{x^{2}-y^{2}}$ pairing corresponds to the
``$d_{xy}$'' pairing of reference 1.  We calculate $U$- and distance
$r$-dependent pair-pair correlations
$P(r)=\frac{1}{2}\langle\Delta_{i}^{\dagger}\Delta_{i+\vec{r}}+\Delta_{i}\Delta^\dagger_{i+\vec{r}}\rangle$,
\begin{figure}[t]
\centerline{\resizebox{3.3in}{!}{\includegraphics[width=3.0in,height=3.1in]{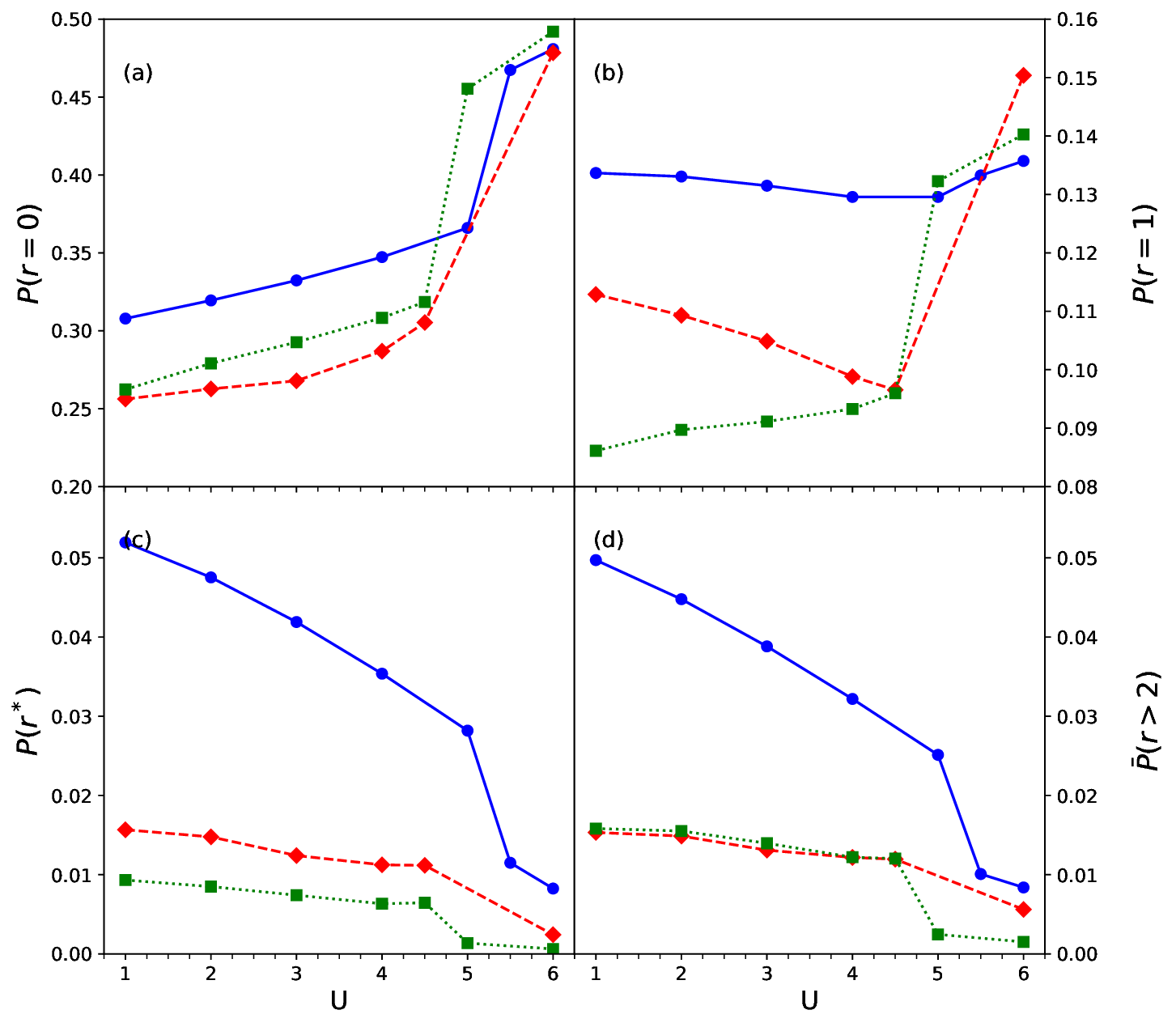}}}
\caption{d$_{x^{2}-y^{2}}$ pair-pair correlations for 4$\times$4 (circles),
6$\times$4 (diamonds), and 6$\times$6 (squares) lattices. (a) $r$=0
(b) $r$=1 (c) $r$=$r^{*}$, and (d) $\bar{P}$ (see text).}
\end{figure}

$\Delta_{i}^{\dagger}=8^{-\frac{1}{2}}\sum_{\nu}g(\nu)(c_{i,\uparrow}^{\dagger}c^\dagger_{i+\nu,\downarrow}-c^\dagger_{i,\downarrow}c^\dagger_{i+\nu,\uparrow})$,
$c_{i,\sigma}^{\dagger}$ creates an electron on site $i$ with spin
$\sigma=\pm\frac{1}{2}$, and the phase factor $g(\nu)$ alternates as
+1, -1, +1, -1 for the four sites ${i+\hat{x}}$, ${i+\hat{y}}$,
${i-\hat{x}}$ and ${i-\hat{y}}$. In addition to $P(r)$ we also
calculated the spin-structure factor $S(\pi,\pi)$ to determine the
occurrence of AFM.

One essential criterion for SC within the model Hamiltonian is simple:
superconducting pair-pair correlations must be enhanced over the $U=0$
values over a minimal range of $U$.  In Fig.~1(a) - (d) we have
plotted $P(0)$, $P(1)$, $P(r^{*})$ and
$\bar{P}=(1/N_{c})\sum_{r>2}P(r)$ against $U$ for all three lattices,
where $r^{*}$ is the next-to-furthest possible separation R between
two lattice points on the finite lattice (r$^{*}$ = 2.24, and 3.16,
and 3.61 in units of lattice constants for the 4$\times$4, 6$\times$4
and 6$\times$6 lattices, respectively), and $N_{c}$ is the total
number of $r>2$ correlations.  Use of r$^{*}$ instead of the furthest
distance avoids finite-size effects associated with the latter on
periodic clusters.

Our results are identical for all three lattices.  There is a sudden
increase or decrease in $P(r)$ at a lattice-specific $U_{c}$,
coincident with an increase in the spin structure factor indicating a
transition to AFM.  $P(0)$ increases with $U$ in all cases while
$P(1)$ increases with $U$ for $U>U_{c}$. Both behaviors are directly
determined by short-range antiferromagnetic spin correlations
unrelated to SC.  In contrast to these both $P(r^{*})$ and $\bar{P}$
decrease continuously with $U$, starting from $U=0$, and the behavior
are qualitatively the same for both $U<U_{c}$ and $U>U_{c}$, clearly
indicating absence of SC.  Most importantly, the order(s) of magnitude
larger magnitudes of $P(0)$ and $P(1)$ over the long-range
counterparts explain why incorrect conclusions are reached within the
CDMFT calculations: momentum-based calculations that place the same
weight on short- versus long-range pair correlations \textit{in a
  small cluster} are apt to give incorrect conclusions. Indeed, that
going beyond the simple dimer Mott-insulator description is essential
for understanding correlated-electron SC in the CTS has been clear
since the discovery of pressure-driven AFM-to-charge
disproportionation-to-SC transitions in
$\beta^{\prime}$-(BEDT-TTF)$_{2}$ICl$_{2}$, which can only be
understood within a $\frac{3}{4}$-band filled description that takes
into account of the charge degrees of freedom internal to the BEDT-TTF
dimers.  Interestingly, experiments on cuprates also find
charge-ordering in the pseudogapped state, perhaps also indicating
that the explanation of SC requires going beyond the simplest Hubbard
Hamiltonian.

{\it Acknowledgments} Work at University of Arizona Tucson was
partially supported by NSF Grant No. NSF-DMR-2301372. Some of the
calculations were performed using high performance computing resources
maintained by the University of Arizona Research Technologies
department and supported by the University of Arizona Technology and
Research Initiative Fund, University Information Technology Services,
and Research, Innovation, and Impact. Computations at Mississippi
State University (MSU) were supported by the MSU High Performance
Computing Collaboratory (HPC$^2$).
\vskip 1pc
\noindent Rupali Jindal and Sumit Mazumdar \\
Department of Physics \\
University of Arizona \\
Tucson, AZ 85721 \\

\noindent R. Torsten Clay \\ Department of Physics \& Astronomy, and
HPC$^2$ Center for Computational Sciences, Mississippi State
University\\ Mississippi State, MS 39762 \\

\end{document}